\documentclass[journal,draftclsnofoot,onecolumn,12pt]{IEEEtran}
\usepackage{amsthm,amssymb,graphicx,multirow,amsmath,color,amsfonts}
\usepackage[update,prepend]{epstopdf}
\usepackage[noadjust]{cite}
\usepackage[latin1]{inputenc}
\usepackage{tikz}
\usepackage{bbm} 
\usepackage{pdfpages}
\usepackage{tabulary}
\usepackage{multirow}
\usepackage{comment}
\usepackage{subfigure}

\def\nb0{{\mathbf{0}}}
\def\nb1{{\mathbf{1}}}

\newtheorem{theorem}{Theorem}

\newtheorem{remark}{Remark}

\allowdisplaybreaks 
\begin{document}
\graphicspath{{./Figures/}}
\title{
Nearest Neighbor and Contact Distance Distribution for Binomial Point Process on Spherical Surfaces}
\author{ Anna Talgat, Mustafa A. Kishk and Mohamed-Slim Alouini 
\thanks{The authors are with King Abdullah University of Science and Technology (KAUST),Thuwal 23955-6900,Saudi Arabia (e-mail:anna.talgat@kaust.edu.sa;mustafa.kishk@kaust.edu.sa; slim.alouini@kaust.edu.sa} 
\vspace{-8mm}}
\maketitle
\begin{abstract}
This letter characterizes the statistics of the contact distance and the nearest neighbor (NN) distance for binomial point processes (BPP) spatially-distributed on spherical surfaces. We consider a setup of $n$ concentric spheres, with each sphere $S_k$ has a radius $r_k$ and $N_k$ points that are uniformly distributed on its surface. For that setup, we obtain the cumulative distribution function (CDF) of the distance to the nearest point from two types of observation points: (i) the observation point is not a part of the point process and located on a concentric sphere with a radius $r_e<r_k\forall k$, which corresponds to the contact distance distribution, and  (ii) the observation point belongs to the point process, which corresponds to the nearest-neighbor (NN) distance distribution.
\end{abstract}

\section{Introduction} \label{sec:intro}
Cellular coverage has become one of the top needs of the modern society due to its importance in various applications such as healthcare, remote education, industry, and much more. For that reason, it is important to ensure cellular coverage all over the globe including remote areas, rural regions, and many other under-served locations. However, due to the lack of infrastructure, majority of these areas receive bad coverage due to lack of incentive for network operators to invest in these locations.   Undoubtedly, CubeSat and Low Earth Orbit (LEO) satellite communications have become the main areas of interest with their tremendous developments and high potentials to achieve global connectivity. The high potential of these communications has motivated many recent works, such as~\cite{number,number0, saeed2019cubesat}, to identify technological advances and highlight open problems in this field. Notably, the progresses in LEO satellite communications are providing a promising solution to the coverage problem in under-served locations~\cite{dang_6g,9042251}. In particular, by deploying satellite gateways in such regions, coverage can be significantly enhanced using satellite communications. This system architecture requires less expenses compared to typical cellular architectures. In particular, it does not require the extension of optical fibers to such remote locations, which is typically needed to provide core-connection to the deployed base stations. This is replaced in the new setup with the wireless link between the gateway and the satellite. 
Recent advances in LEO satellite research have encouraged various companies to invest in launching large numbers of LEO satellites to ensure low latency communication, for example, SpaceX has received permission to create a constellation of 4425 LEO satellites~\cite{number1}. 

The spatial distribution of the LEO satellite strongly affects the performance of the satellite communication systems. In this paper, we propose to model the locations of the satellites using tools from stochastic geometry. Stochastic geometry is one of the mathematical tools that enable tractable modeling of various types of wireless networks and analyzing their properties~\cite{book}. We develop a new tractable approach where we model the locations of the LEO satellites as a BPP on a sphere. The developed framework is essential for studying the performance of the LEO satellite communication system. However, it is first needed to understand the fundamental characteristics of the distances emerging from this point process, which is the main contribution of this paper.

\subsection{Related work}
 Characterizing the statistics of the distances between various components of the wireless networks is essential for rigorous performance analysis. Relevant literature has mainly focused on spatial point processes on a 2D plane. For instance, authors in~\cite{7922493,number4,8023870} characterize the CDFs of contact and nearest-neighbor distances for Poisson hole processes and Poisson cluster process, respectively. Statistical research on point processes on the sphere could date back to the 1970s, such as a study of random sets on the sphere by Mile's~\cite{number2}. Statistical methods that are developed for analyzing a distribution of points on a spherical region, including modeling and estimating techniques for a specified model, are studied in the recent work~\cite{number3}. However, statistical analysis for contact and nearest neighbor distances for point processes on spherical surfaces are still surprisingly underdeveloped. It is important to point out to the difference between the analysis of point processes in 3-dimensional (3D) plane, which is relatively well-understood part of literature~\cite{number5}, and the analysis of point processes solely distributed on a spherical surface. Only very recently, during writing this paper, a new work has tackled this problem while modeling the location of the satellites as a BPP on a sphere~\cite{9079921}. The main differences between this paper and~\cite{9079921} is: (i) we study a more general model where points are randomly located on multiple concentric spheres, which resembles the scenario of having the satellites at multiple altitudes as is the case in majority of announced constellation by key-companies in this area, and (ii) in addition to deriving the contact distance distribution, we also derive the nearest neighbor distribution for a satellite on the $k$th sphere, which is an important metric that has its own value for studying routing between LEO satellites. A deterministic version of the setup considered in this paper was studied in~\cite{number6,number8} with the objective of optimizing the LEO satellite constellations..

\subsection{Contributions}
The main contributions of this work are as follows.
First, we  model $n$  concentric spheres with $N_k$ points uniformly distributed on each sphere $\forall k$. Then we use tools from stochastic geometry to provide a new tractable model for studying distance distributions in satellite networks located on spherical surfaces. In particular, we model the location of points as a spherical BPP to study the distribution of nearest-neighbor distance for two different locations of observation point which are (i) observation point is not a part of the point process and located on the Earth, and  (ii) observation point is a part of the point process and located on $k^{th}$ sphere. Closed-form expressions for the distance distributions are derived and verified using Monte-Carlo simulations. Finally, with the assistance of numerical results, various system-level insights are drawn and discussed. 

\section{System Model} \label{sec:SysMod}

\begin{figure}
\centering
  \includegraphics[width=0.67\columnwidth]{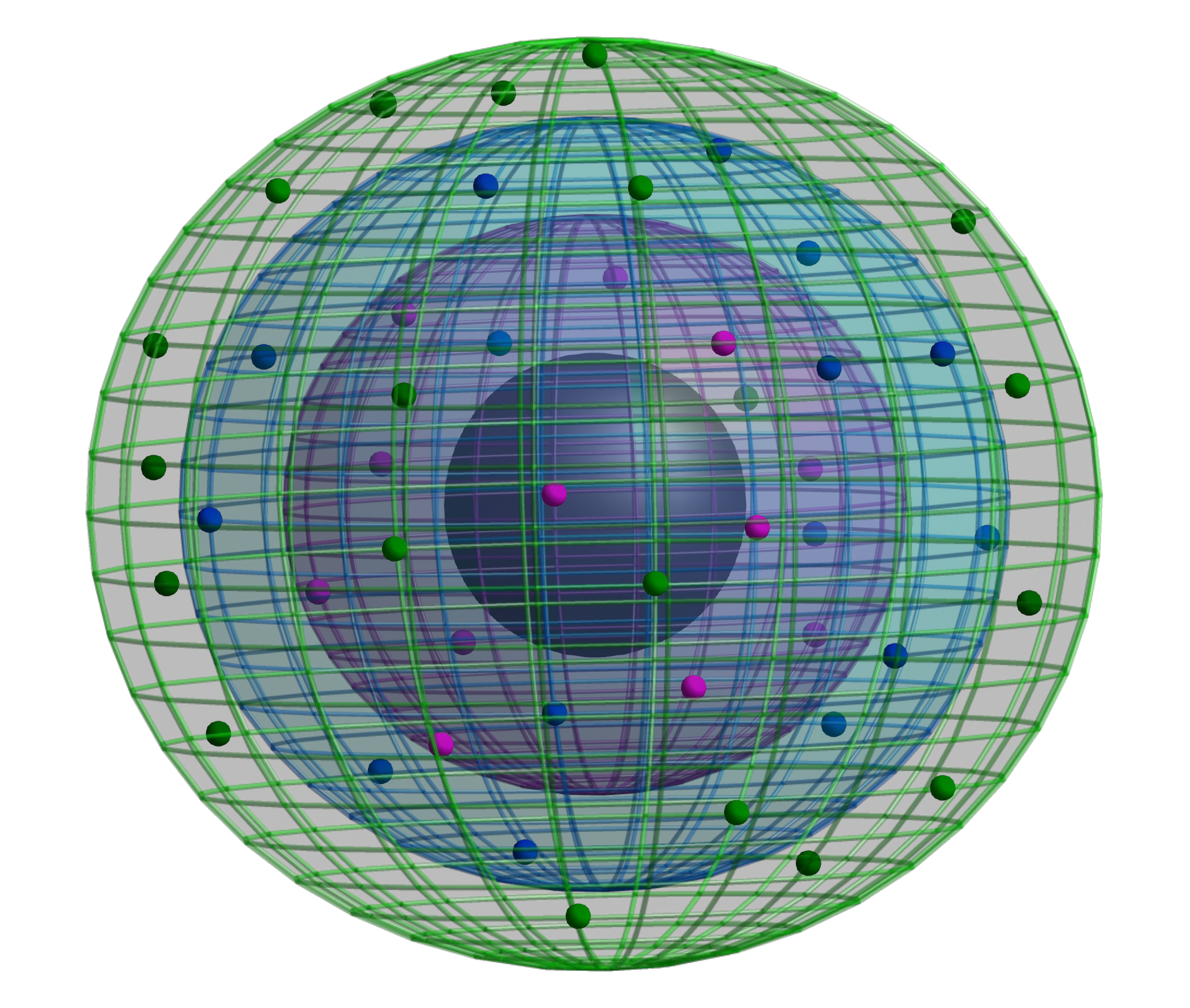}
  \caption{ System model for $n$ level of spheres concentric with the Earth. }
   \label{fig:1}
 \end{figure}
As stated above, the analysis in this paper is motivated by the recent advances in the area of LEO satellite communication systems. Hence, our objective is to provide a model that captures two kinds of communication links: (i) links between gateways on the earth and LEO satellites, and (ii) inter-satellite links between LEO satellites. For the former, it is important to derive the distribution of the distance between a point on the earth and its nearest LEO satellite. For the latter, in order to study backhaul communication between LEO satellites, it is important to derive the distribution of the distance between a given LEO satellite and its nearest neighbor.  We  model this system setup as in  Fig.~\ref{fig:1} which represents a random LEO constellation in which a set of satellites is distributed on set of spheres  according to a BPP.  In particular, 
we consider a system composed of $n$ concentric spheres, denoted by $S_k\subset\mathbb{R}^3$, $\forall 1\leq k\leq n$. On each sphere, a point process ${\Phi}_{k}$ composed of $N_k$ points are uniformly distributed. Each sphere is defined by the altitude $a_k$ (altitude of $k^{th}$ sphere from the surface of Earth) and the radius $r_k=r_e+a_k$, where $r_e$ is the radius of the earth. Hence, the considered point process is defined as
 ${\Phi}=\bigcup_{k=1}^{n}{{\Phi}_{k}}$ on $\bigcup_{k=1}^{n}{S_k}$.  We denote its corresponding counting measure by $N$, such that $N(\mathcal{A})$ denotes the number of points in ${\Phi}$ falling in the region $\mathcal{A} \subseteq \bigcup_{k=1}^{n}S_k$. For each BPP ${\Phi}_{k}$, fixed number $N_k$ of points are independently and uniformly distributed on a sphere $S_k$ defined as

$S_k\overset{\Delta}{=}$\{($r_k$, $\varphi$, $\theta$): $r_k=r_e+a_k$, $0\leq \varphi \leq \pi$, $0\leq\theta <2\pi$\}, where the ($r_k$, $\varphi$, $\theta$) represent the spherical coordinates in $\mathbb{R}^3$. The nearest neighbor or contact distance (depending on the definition of the observation point) is the distance from the observation point to the nearest point in $\Phi$ and is given by $D$. The corresponding distribution  $F_{D}(d)  \overset{\Delta}{=} \mathbb{P}(D< d)$ is the nearest neighbor or contact distance distribution function.

\begin{table*}
\centering
\caption{SUMMARY OF NOTATION}
\label{tab:example}
\centering
\begin{tabular}{c|c}
    \hline
    \textbf{Notation } &  \textbf{Description}\\
    \hline
    \hline
    $\Phi_{k}$; $N_k$  &  BPP modeling the locations of point; number of point on $k^{th}$ sphere \\
    \hline
    $S_k$; $r_k$; $a_k$   & $k^{th}$ sphere with radius $r_k$ and altitude $a_k$ to the surface of the Earth.  \\
    \hline
     $h(d,k,i)$   &   height of spherical cap, $(k,i)$ represent the spheres where nearest distance and observation point are located respectively. \\
   \hline
    $h_{\rm max}(k,i)$   & height of spherical cap formed by the maximum distance $d_{\rm max}(k,i)$.  \\
    \hline
    \hline
 \end{tabular}
\end{table*}
\begin{figure}
\centering
\includegraphics[width=0.56\columnwidth]{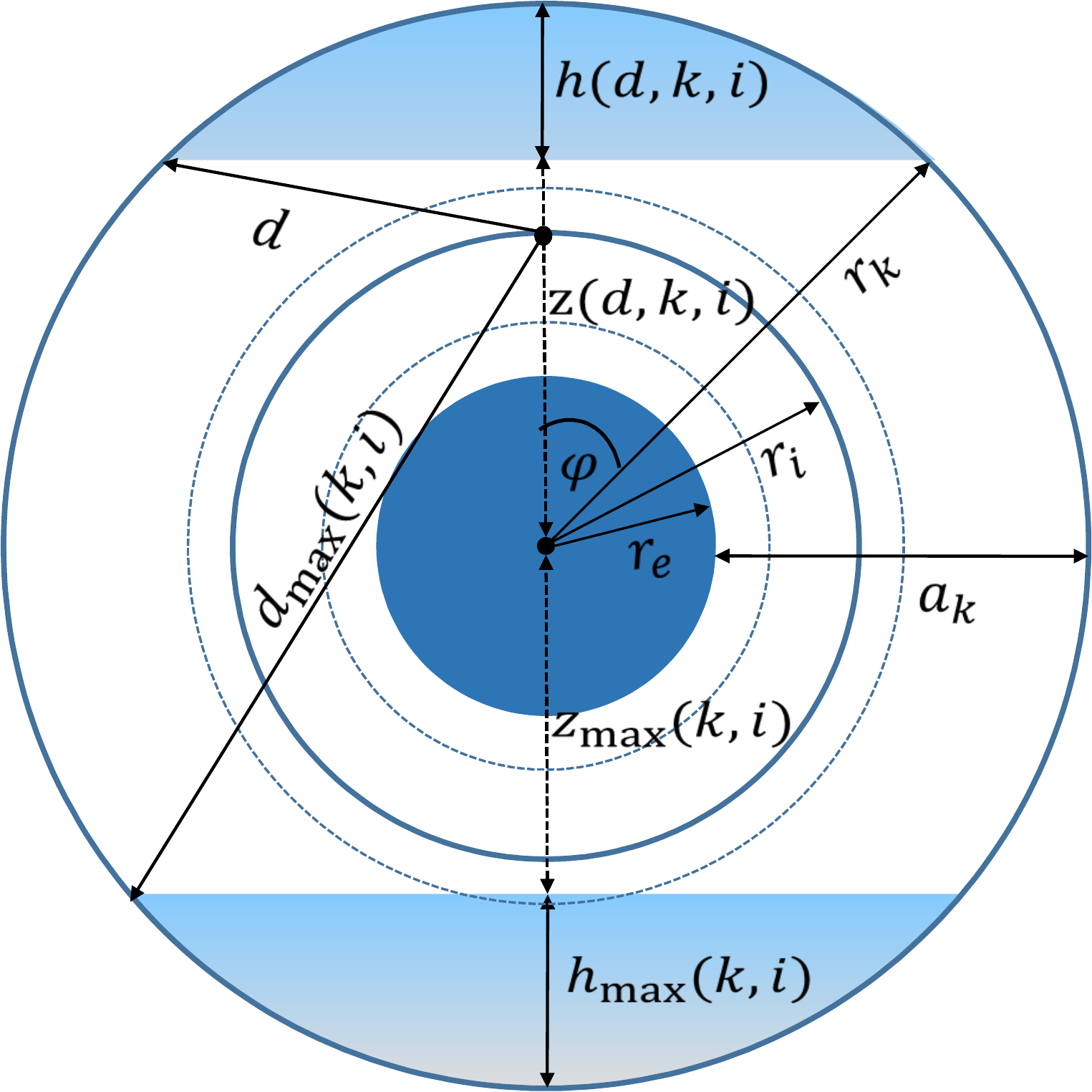}
\caption{Observation point located on the $i^{th}$ sphere.}
\label{fig:2}
\end{figure}
\subsection{Scenario-1 Description} 
The observation point is located on the Earth. The corresponding distribution is
$$F_{D}(d)\overset{\Delta}{=}  \mathbb{P}(D< d)=1-\prod_{k=1}^{n}  \mathbb{P}(D_k \geq d),$$ where $ \mathbb{P}(D_k \geq d)=\bar{F}_{D_k}(d)=1-{F}_{D_k}(d)$ is the complementary cumulative distribution function (CCDF) of the contact distance $D_k$ from the observation point to the nearest point on $k^{th}$ sphere.

By definition, we know that if $d<a_k$ then $F_{D_k}(d)=0$. For $d>a_k$, ${F}_{D_k}(d)$ is the probability that the number of points on a given spherical cap $\mathcal{A}_k$ at height $h(d,k,0)$ is greater than zero:
$$F_{D_k}(d)\overset{\Delta}{=}  \mathbb{P}(D_k< d)= \mathbb{P}(N(\mathcal{A}_k)>0).$$
Hence, the CCDF of $D_k$ can be computed as follows.
$$ \mathbb{P}(D_k \geq d)= \mathbb{P}(N(\mathcal{A}_k)=0)=[ \mathbb{P}(z_k< z(d,k,0))]^{N_k}, $$ where $z_k=r_k \cos{\varphi}$, and $z(d,k,0)=r_k-h(d,k,0)$.
With Pythagoras' theorem,  we can easily derive the expression of $h(d,k,0)$. Assuming that the communication between any point on the earth and an LEO satellite requires a Line-of-Sight (LoS), the maximum distance, $d_{\rm max}(k,0)$, that can be taken from the observation point  also forms a spherical cap $\mathcal{A}_{\rm max,k}$ with height $h_{\rm max}(k,0)$. When the number of points in $\mathcal{A}_{\rm max,k}$ is zero, it means that there are no points in $S_k$ that have an LoS with the observation point. Hence, for that scenario, we assume that $D_k=\infty$. As a result, the CCDF of $D_k$ for $d>d_{\rm max}(k,0)$ is 
$$ \mathbb{P}(D_k \geq d)= \mathbb{P}(N(\mathcal{A}_{\rm max,k})=0)=[ \mathbb{P}(z_k< z_{\rm max}(k,0))]^{N_k}, $$ where  $z_{\rm max}(k,0)=r_e$.
Combining all the conditions which are  $d<a_k$, $a_k<d<d_{\rm max}(k,0)$ and $d>d_{\rm max}(k,0)$, we can derive $F_{D_k}(d)$  $\forall k$. 
\subsection{Scenario-2 Description}
The observation point is located on the $S_i$, and the point is part of the point process. So, the corresponding distribution is
{\small{$$F_{D}(d)=1-\big[\prod_{k=1}^{i-1} \mathbb{P}(D_{k,i} \geq d)\big]\big[ \mathbb{P}(D_i \geq d)\big]\big[\prod_{k=i+1}^{n} \mathbb{P}(D_{k,i}\geq d)\big],$$}}
where $D_{k,i}$ is the distance between the observation point and the nearest point on $S_k$, and $D_i$ is the distance between the observation point and the nearest point on the same sphere $S_i$. Here, the CCDFs correspond to NN distance distribution for  {(a)} below the $i^{th}$ sphere, {(b)} on the $i^{th}$ sphere and   {(c)} above the $i^{th}$ sphere  respectively.
 Fig.~\ref{fig:2} shows the communication link  for the case {(c)} while the observation point is located on the sphere $S_i$. The shaded parts represent the spherical caps $\mathcal{A}_k$ and $\mathcal{A}_{\rm max,k}$  formed on  $S_k$ with height  $h(d,k,i)$ and $h_{\rm max}(k,i)$, respectively.  
We follow the same procedure as Scenario-1 to derive the complete CDF for each case where conditions are $|a_k-a_i|<d,$ $|a_k-a_i|<d<d_{\rm max}(k,i)$  and $d> d_{\rm max}(k,i)$ for {(a)} and {(c)} and $d<d_{\rm max}(k,i)$ and $d>d_{\rm max}(k,i)$ for case {(b)}. 
Also, we get  $h(d,k,i)$, $h_{\rm max}(i,i)$ and $d_{\rm max}(k,i)$ for each case separately by using Pythagoras' theorem.

\section{Distance Distribution} 
In this section, we determine the distribution of the nearest distance from a specified observation point for a general BPP. 

\begin{theorem}[Scenario-1: Contact distance distribution] \label{thm:1} 
The CDF of the distance D from the observation point arbitrarily located on the surface of Earth to  nearest  satellite in the constellation  is given by
\begin{align}
  F_{D}(d)  \overset{\Delta}{=}  \mathbb{P}(D< d)= 1- \prod_{k=1}^{n}  \mathbb{P}(D_k \geq d),
\end{align}
where the CCDF of $D_k$ is\\

$ \mathbb{P}( D_k\geq d)=$
{\small
\[\left\{\begin{array}{ll}
1,&d<a_k\\
\big[1-\frac{1}{\pi}\arccos{(1-\frac{{d^2-a_{k}^2}}{2r_er_k})})\big]^{N_k}, & a_k\leq d\leq d_{\rm max}(k,0)\\
\big[{1-\frac{1}{\pi}\arccos(\frac{r_e}{r_k})}\big]^{N_k},& d>d_{\rm max}(k,0),
\end{array}\right.\]}
where $d_{\rm max}(k,0)=\sqrt{2r_ea_k+a_{k}^2}$.
\begin{IEEEproof}
See Appendix~\ref{app:1}.
\end{IEEEproof}
\end{theorem}

\begin{theorem}[Scenario-2: Nearest neighbor distance distribution]
The CDF of the distance D from the observation point chosen randomly from BPP on  $i^{th}$ sphere to its  nearest  satellite in the constellation  is given by
\label{thm:2}
\begin{align}
   F_{D}(d)  \overset{\Delta}{=} \mathbb{P}(D<d)=1-\prod_{\substack{k=1}}^{n}  \mathbb{P}(D_k \geq d),  
\end{align}
where the CCDFs of $D_k$ are described below.\\

For $k=i$, 
{\small
\begin{equation*}
 \mathbb{P}(D_i\geq d) =
\left\{
\begin{array}{rl}
\big[1- \frac{1}{\pi}\arccos{(1-\frac{{d^2}}{ 2r_{i}^2}})\big]^{N_{i}-1}, &  d<d_{\rm max}(i,i) \\
\big[1- \frac{1}{\pi}\arccos{(1-\frac{{2r_e^2}}{r_{i}^2}})\big]^{N_{i}-1},& d > d_{\rm max}(i,i),
\end{array}\right.
\end{equation*}} \\
where $d_{\rm max}(i,i)=2\sqrt{r_{i}^2-r_{e}^2}$.\\

For $k\neq i$, $ \mathbb{P}(D_k\geq d) =$
\small
\begin{equation*}
\left\{
\begin{array}{ll}
1,\hspace{6.1cm} d<|a_k-a_i| \\
\big[1- \frac{1}{\pi}\arccos{(1-\frac{{d^2-(a_{i}-a_{k})^2}}{ 2r_i r_k}})\big]^{N_k}, \\ \hspace{4.7cm}  |a_k-a_i|\leq d \leq d_{\rm max}(k,i)\\
\big[1- \frac{1}{\pi}\arccos{(1-\frac{{(r_i+r_k)^2-d^2_{\rm max}(k,i)}}{ 2r_i r_k}})\big]^{N_k},\hspace{0.5cm} d > d_{\rm max}(k,i),
\end{array}\right.
\end{equation*}
where ${d_{\rm max}(k,i)=\sqrt{r_k^2-r_e^2}+\sqrt{r_i^2-r_e^2}}$.
\begin{IEEEproof}
See Appendix~\ref{app:2}.
\end{IEEEproof}
\end{theorem}

\begin{remark}
The results in Theorems~1 and 2 show the influence of the value of $N_k$ on the distribution of both the contact distance and the nearest neighbor. In particular, we notice that the value $\mathbb{P}(D_k\geq d)$ in both theorems includes a value raised to the power of $N_k$, where this value is less than 1. This, in turns, implies that $\mathbb{P}(D_k\geq d)$ is a decreasing function of $N_k$. This intuitive insight, supported by our analytical expressions, can be used to select the proper values of $N_k$ to achieve a specific level of system performance.
\end{remark}

\section{Numerical Results} 

\begin{figure}
\includegraphics[width=0.9\columnwidth]{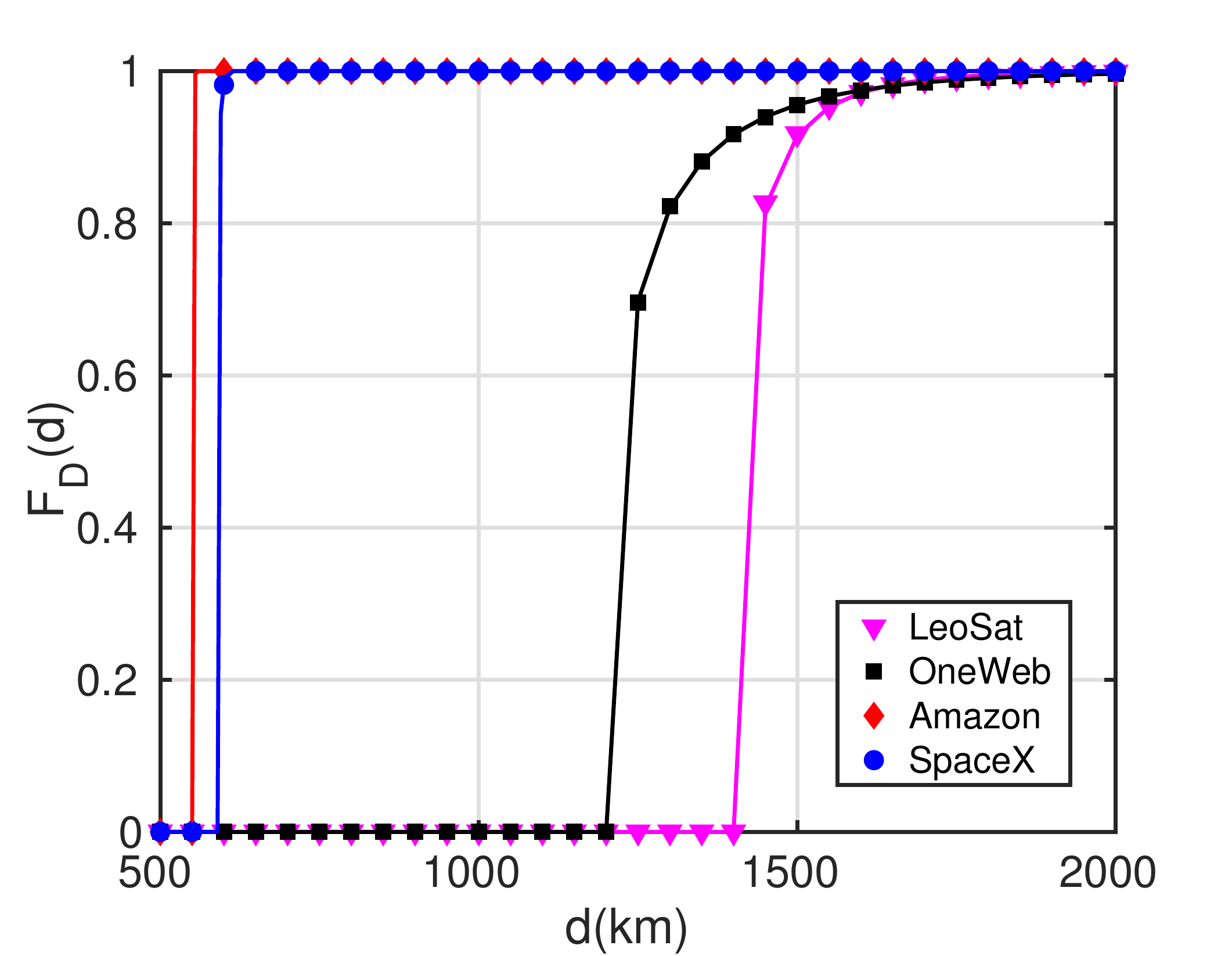}
\caption{Practical case: CDF of contact distance  for different companies: $a_{k_{\rm LeoSat}}=$[1400] and   $N_{k_{\rm LeoSat}}$=[100]; 
$a_{k_{\rm OneWeb}}=$[1200] and   $N_{k_{\rm OneWeb}}$=[74]; $a_{k_{\rm Amazon}}=$[590 610 630] and  $N_{k_{\rm Amazon}}$=[784 1296 1156]; $a_{k_{\rm SpaceX}}=$[550 1110 1130 1275 1325] and $N_{k_{\rm SpaceX}}$=[1584 1600 400 374 450].}
\label{fig:5}
\end{figure}

\begin{figure}
\centering
\includegraphics[width=0.9\columnwidth]{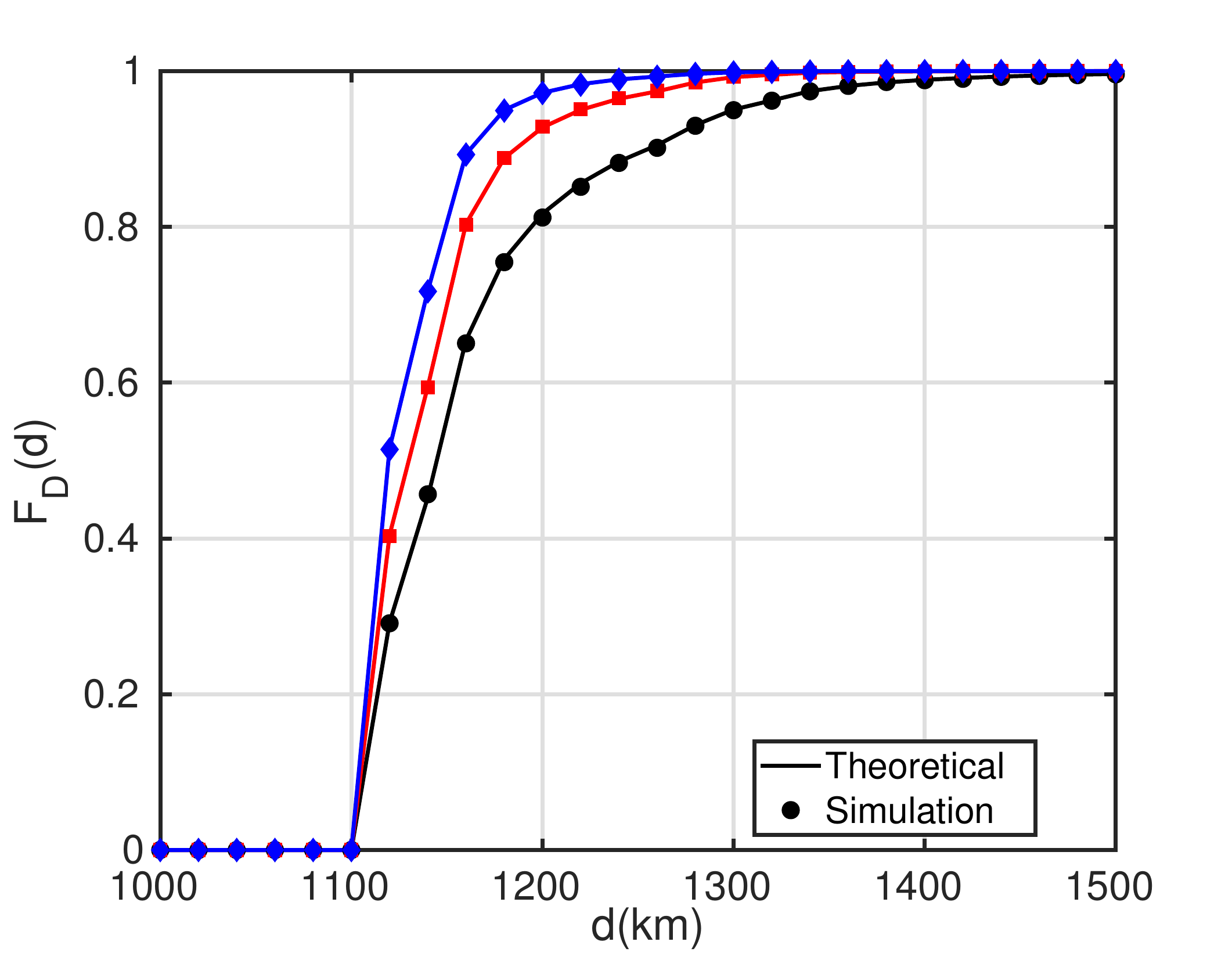}
\caption{Scenario-1: CDF of contact distance  for different number of multi-level spheres with values: 
$a_{k_{\rm circle}}=$[1110 1150 1275 1325] and   $N_{k_{\rm circle}}$=[50 40 25 15]; $a_{k_{\rm square}}=$[1110 1150 1275 1325 1500 1700] and $N_{k_{\rm square}}$=[75 65 55 45 25 15]; $a_{k_{\rm diamond}}=$[1110 1150 1275 1325] and  $N_{k_{\rm diamond}}$=[105 85 60 35].}
\label{fig:3}
\end{figure}

In this section, we provide numerical results for the derived distance distributions. As shown in Fig.~\ref{fig:3} and Fig.~\ref{fig:4}, Theorems \ref{thm:1} and \ref{thm:2}  are perfectly matching with simulation, which affirms the accuracy of our analysis. 

 Fig.~\ref{fig:5} represents the LEO satellite constellation for four different companies.
  The solid lines indicate the CDF values computed using theoretical expressions, while the markers represent the simulation results.
 
In Fig.~\ref{fig:3}, we plot the CDF of the contact distance for three different system setups, as described in the caption. We observe that the number of satellites on each sphere, when the altitudes are fixed, have a noticeable influence on the distribution of the contact distance. By comparing the scenarios represented by the diamond and circle-shaped markers, we notice that for similar altitudes of the spheres, the scenario that has larger number of satellites has larger value of CDF of the distance. This kind of system level insights can be useful for system architecture designers to select the optimum altitudes and number of satellites to maximize the system performance.


In Fig.~\ref{fig:4}, we plot the CDF of the distance to the nearest neighbor distance as the observation point moves from $S_1$ to $S_n$ where $n=4$. We notice that the distance to the nearest neighbor gets higher as the observation point moves from inner spheres to outer spheres. This insight is of high importance in the context of inter-satellite multi-hop communication since it implies the more challenging routing options for the satellites existing at the highest altitudes.

\section{Conclusion}

In this letter, we presented a stochastic geometry framework to model the spatial distribution of LEO satellite communication systems. For that setup, we derived the exact analytical expressions for the CDFs of the nearest neighbor and the contact distance where fixed numbers of points are independently and uniformly distributed on a number of concentric spheres. The provided setup can be used in various applications such as (i) studying the coverage probability of the LEO-aided communication networks and (ii) studying the routing among LEO satellites.
\appendix

\subsection{Proof of Theorem~\ref{thm:1}}\label{app:1}

The proof of Theorem 1 and 2 follow the same steps.
\begin{itemize}
    \item If $d< a_k$, we have $ \mathbb{P}(D_k \geq d)=1$  $\forall k $.\\
    \item If $a_k< d<d_{\rm max}(k,0)$, then  we have the contact distance distribution,
\small{
\begin{align*}
 \mathbb{P}&(D_{k}\geq d)= \mathbb{P}(N(\mathcal{A}_{k})=0)\\
   &= [ \mathbb{P}(z_k <z(d,k,0))] ^{N_k} \\
   &= [ \mathbb{P}(r_k \cos{\varphi} < z(d,k,0))]^{N_k}\\
   &=[ \mathbb{P}(\varphi>\arccos{\frac{z(d,k,0)}{r_k}})+ \mathbb{P}(\varphi<-\arccos{\frac {z(d,k,0)}{r_k}})]^{N_k}\\
   &=[ 1-\frac{1}{\pi }\arccos(\frac{z(d,k,0)}{r_k})] ^{N_k},
\end{align*}}\noindent where $h(d,k,0)=\frac{d^2-a_k^2}{2r_e}$ and $d_{\rm max}(k,0)=\sqrt{2r_ea_k+a_{k}^2}$  are easily derived  by Pythagoras' theorem and  $z(d,k,0)=r_k-h(d,k,0)$.\\
\item If $d> d_{\rm max}(k,0)$, skipping all the intermediate steps we get
\small{\begin{align*}
 \mathbb{P}D_{k}\geq d)&= \mathbb{P}(N(\mathcal{A}_{\rm max,k})=0)= [ \mathbb{P}(z_k <z_{\rm max}(k,0))] ^{N_k}\\
  &=[{1 -\frac{1}{\pi} \arccos(\frac{z_{\rm max}(k,0)}{r_k})}] ^{N_k},
\end{align*}}
where  $z_{\rm max}(k,0)=r_e $ $ \forall k $.
\end{itemize} 
This concludes the proof.

\begin{figure}
\includegraphics[width=0.9\columnwidth]{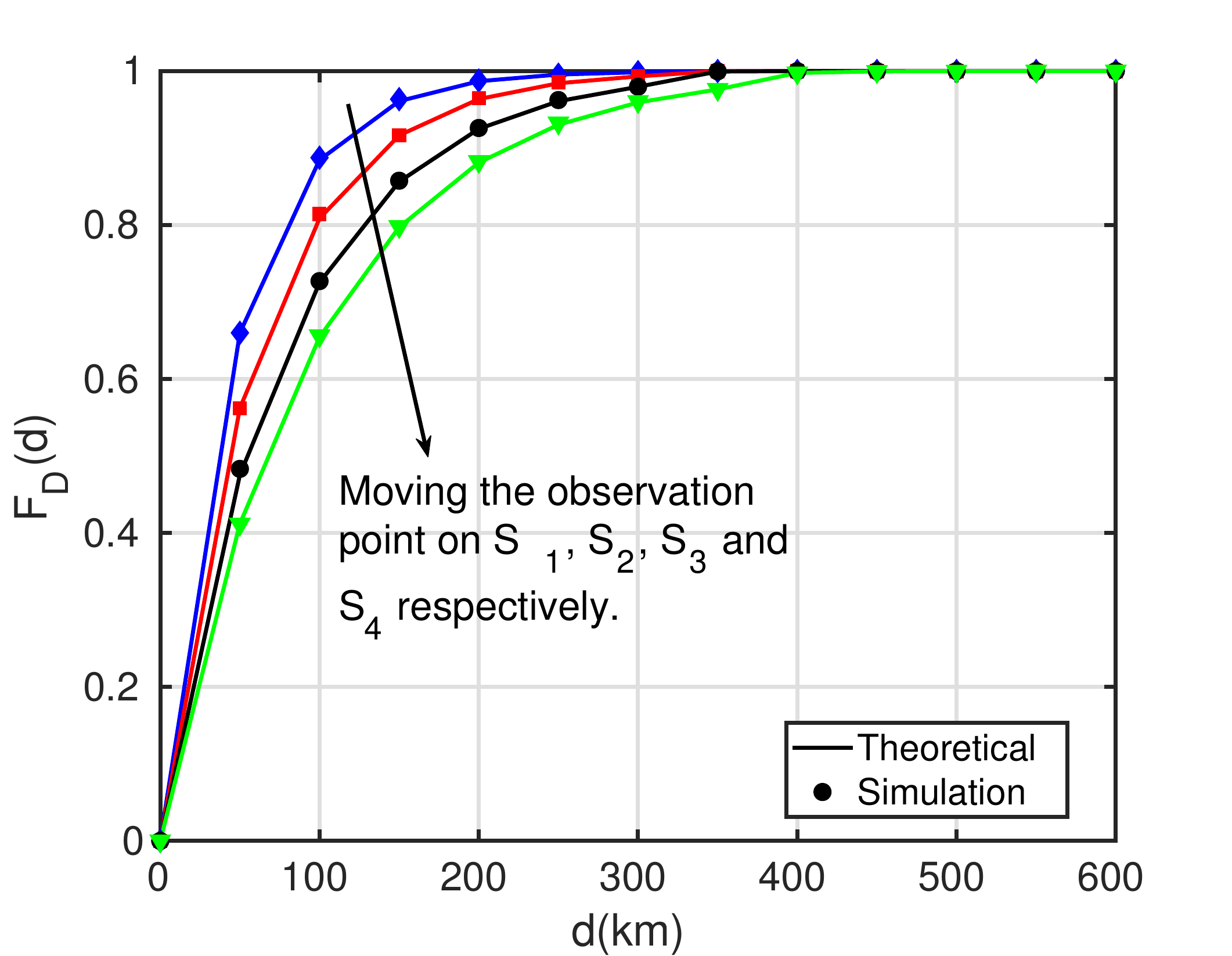}
\caption{Scenario-2: CDF of the nearest-neighbor distance for a setup composed of 4 spheres as follows: $S_1$= [1000   500], $S_2$= [1325  400], $S_3$= [1625  325] and  $S_4$= [2000  280], where $S_k=[a_k  N_k]$.}
\label{fig:4}
\end{figure}

\subsection{Proof of Theorem~\ref{thm:2}}\label{app:2}

For $S_k$ and $ k\neq i$:  
\begin{itemize}
\item If $d<|a_k-a_i|$, we have $ \mathbb{P}(D_k \geq d)=1$.\\
\item If $|a_k-a_i|< d<d_{\rm max}(k,i)$,
\small{
\begin{align*}
 \mathbb{P}(D_{k}\geq d)&= \mathbb{P}(N(\mathcal{A}_{k})=0)=[ \mathbb{P}(z_k<z(d,k,i))]^{N_k} \\
   &=[1-\frac{1}{\pi}\arccos(\frac{z(d,k,i)}{r_k})]^{N_k},
\end{align*}}\noindent  where ${h(d,k,i)=\frac{d^2-(a_{i}-a_{k})^2}{2r_i}}$, ${z(d,k,i)=r_k-h(d,k,i)}$, and ${d_{\rm max}(k,i)=\sqrt{r_k^2-r_e^2}+\sqrt{r_i^2-r_e^2}}$.\\
\item If $d> d_{\rm max}(k,i)$, 
\small{\begin{align*}
 \mathbb{P}(D_{k}\geq d)&= \mathbb{P}(N(\mathcal{A}_{\rm max,k})=0)= [ \mathbb{P}(z_k<z_{\rm max}(k,i))]^{N_k}\\
   &=[1-\frac{1}{\pi} \arccos(\frac{z_{\rm max}(k,i)}{r_k})]^{N_k},
\end{align*}}\noindent where ${h_{\rm max}(k,i)=\frac{(r_i+r_k)^2-d^2_{\rm max}(k,i)}{2r_i}}$ and $z_{\rm max}(k,i)=r_{k}-h_{\rm max}(k,i)$.\\
\end{itemize}

For $S_k$ and  $k=i $, the observation point is part of the point process ${\Phi}_{i}$, the remaining point process becomes a spherical BPP with $N_{i}-1$ points.  
\begin{itemize}
\item If $d<d_{\rm max}(i,i)$, then 
\begin{align*}
 \mathbb{P}(D_{i}\geq d)&=  \mathbb{P}(N(\mathcal{A}_{i})=0)=[ \mathbb{P}(z_i <z(d,i,i))]^{N_i-1}\\
             &=[1-\frac{1}{\pi}\arccos(\frac{z(d,i,i)}{r_i})]^{N_i-1},
\end{align*}
where $h(d,i,i)=\frac{d^2}{2 r_i}$, $d_{\rm max}(i,i)=2\sqrt{r_{i}^2-r_{e}^2}$ and 
with $z(d,i,i)=r_i-h(d,i,i)$. \\
\item  If $d> d_{\rm max}(i,i)$, then 
\begin{align*}
 \mathbb{P}(D_{i}\geq d)&=  \mathbb{P}(N(\mathcal{A}_{\rm max,i})=0)\\ & =[ \mathbb{P}(z_i<z_{\rm max}(i,i))]^{N_{i}-1} \\
   &=[1-\frac{1}{\pi}\arccos(\frac{z_{\rm max}(i,i)}{r_i})]^{N_{i}-1},
\end{align*}
where $h_{\rm max}(i,i)=\frac{2r_e^2}{r_i}$ and  $z_{\rm max}(i,i)=r_{i}-h_{\rm max}(i,i)$.
\end{itemize}
This concludes the proof.

\bibliographystyle{IEEEtran}
\bibliography{Draft_v0.1.bbl}

\begin{thebibliography}{10}
\providecommand{\url}[1]{#1}
\csname url@samestyle\endcsname
\providecommand{\newblock}{\relax}
\providecommand{\bibinfo}[2]{#2}
\providecommand{\BIBentrySTDinterwordspacing}{\spaceskip=0pt\relax}
\providecommand{\BIBentryALTinterwordstretchfactor}{4}
\providecommand{\BIBentryALTinterwordspacing}{\spaceskip=\fontdimen2\font plus
\BIBentryALTinterwordstretchfactor\fontdimen3\font minus
  \fontdimen4\font\relax}
\providecommand{\BIBforeignlanguage}[2]{{%
\expandafter\ifx\csname l@#1\endcsname\relax
\typeout{** WARNING: IEEEtran.bst: No hyphenation pattern has been}%
\typeout{** loaded for the language `#1'. Using the pattern for}%
\typeout{** the default language instead.}%
\else
\language=\csname l@#1\endcsname
\fi
#2}}
\providecommand{\BIBdecl}{\relax}
\BIBdecl

\bibitem{number}
B.~Di, L.~Song, Y.~Li, and H.~V. Poor, ``Ultra-dense {LEO}: Integration of
  satellite access networks into {5G} and beyond,'' \emph{IEEE Wireless
  Communications}, vol.~26, no.~2, pp. 62--69, 2019.

\bibitem{number0}
O.~Kodheli, E.~Lagunas, N.~Maturo, S.~K. Sharma, B.~Shankar, J.~F.~M. Montoya,
  J.~C.~M. Duncan, D.~Spano, S.~Chatzinotas, S.~Kisseleff, J.~Querol, L.~Lei,
  T.~X. Vu, and G.~Goussetis, ``Satellite communications in the new space era:
  A survey and future challenges,'' available online: arxiv.org/abs/2002.08811.

\bibitem{saeed2019cubesat}
N.~{Saeed}, A.~{Elzanaty}, H.~{Almorad}, H.~{Dahrouj}, T.~Y. {Al-Naffouri}, and
  M.~{Alouini}, ``Cubesat communications: Recent advances and future
  challenges,'' {\em IEEE Communications Surveys Tutorials}, to appear.

\bibitem{dang_6g}
S.~{Dang}, O.~{Amin}, B.~{Shihada}, and M.-S. {Alouini}, ``What should {6G}
  be?'' \emph{Nature Electronics}, vol.~3, no.~1, pp. 20--29, Jan. 2020.

\bibitem{9042251}
E.~{Yaacoub} and M.~{Alouini}, ``A key {6G} challenge and
  opportunity-connecting the base of the pyramid: A survey on rural
  connectivity,'' \emph{Proceedings of the IEEE}, vol. 108, no.~4, pp.
  533--582, 2020.

\bibitem{number1}
M.~Handley, ``Delay is not an option: Low latency routing in space,''
  \emph{HotNets '18: The 17th ACM workshop on Hot Topics in Networks}, vol.~33,
  no.~2, pp. 85--91, 2018.

\bibitem{book}
M.~Haenggi, \emph{Stochastic Geometry for Wireless Networks}.\hskip 1em plus
  0.5em minus 0.4em\relax Cambridge University Press, 2012.

\bibitem{7922493}
M.~A. {Kishk} and H.~S. {Dhillon}, ``Tight lower bounds on the contact distance
  distribution in {P}oisson hole process,'' \emph{IEEE Wireless Communications
  Letters}, vol.~6, no.~4, pp. 454--457, 2017.

\bibitem{number4}
M.~Afshang, C.~Saha, and H.~S. Dhillon, ``Nearest-neighbor and contact distance
  distributions for {T}homas cluster process,'' \emph{IEEE Wireless
  Communications Letters}, vol.~6, no.~1, pp. 130--133, 2016.

\bibitem{8023870}
M.~{Afshang}, C.~{Saha}, and H.~S. {Dhillon}, ``Nearest-neighbor and contact
  distance distributions for {M}atern cluster process,'' \emph{IEEE
  Communications Letters}, vol.~21, no.~12, pp. 2686--2689, 2017.

\bibitem{number2}
R.~E. Miles, ``Random points, sets and tessellations on the surface a sphere,''
  \emph{Sankhya: The Indian Journal of Statistics}, vol.~33, no.~2, pp.
  145--174, 1971.

\bibitem{number3}
T.~Lawrence, A.~Baddeley, R.~K. Milne, and G.~Nair, ``Point pattern analysis on
  a region of a sphere,'' \emph{STAT}, vol.~5, no.~1, pp. 144--157, 2016.

\bibitem{number5}
M.~Haenggi, ``On distances in uniformly random networks,'' \emph{IEEE
  Transaction on Information Theory}, vol.~51, no.~10, pp. 3584--3586, 2005.

\bibitem{9079921}
N.~{Okati}, T.~{Riihonen}, D.~{Korpi}, I.~{Angervuori}, and R.~{Wichman},
  ``Downlink coverage and rate analysis of low earth orbit satellite
  constellations using stochastic geometry,'' {\em IEEE Transactions on
  Communications}, to appear.

\bibitem{number6}
B.~Soret, I.~Leyva-Mayorga, and P.~Popovski, ``Inter-plane satellite matching
  in dense {LEO} constellations,'' \emph{IEEE Transaction on Information
  Theory}, vol.~1, no.~10, pp. 3584--3586, 2019.

\bibitem{number8}
Z.~Qu, G.~Zhang, H.~Cao, and J.~Xie, ``{LEO} satellite constellation for
  internet of things,'' \emph{IEEE Access}, vol.~5, pp. 18\,391 -- 18\,401,
  2017.

\end{thebibliography}
\end{document}